\documentclass[prd,showpacs,twocolumn,preprintnumbers
]{revtex4}

\usepackage{amsmath} \usepackage{graphicx}

\newcommand{\be}{\begin{equation}} \newcommand{\ee}{\end{equation}}
\newcommand{\bel}{\begin{align}} \newcommand{\eel}{\end{align}}
\newcommand{\bea}{\begin{eqnarray}} \newcommand{\eea}{\end{eqnarray}}
 
\newcommand{\mn}{{\mu\nu}} \newcommand{\lb}{\label}
\newcommand{\Lef}{\left(} \newcommand{\Rig}{\right)}
 \newcommand{\cF}{{\cal F}}
\newcommand{\cG}{{\cal G}} \newcommand{\cH}{{\cal H}}
\newcommand{\cR}{{\cal R}} \newcommand{\e}{{\rm e}}
\newcommand{\2}{\Gamma} \newcommand{\3}{\Pi_\Psi}
\newcommand{\4}{\Pi_\Gamma} \newcommand{\1}{\Psi}
\newcommand{\x}{p} \newcommand{\y}{q} \newcommand{\z}{r}
\newcommand{\w}{s}
\newcommand{\duF}{\tilde{F}}

\begin{document}

\preprint{DTP-MSU/04-20} \preprint{hep-th/0412334}
\preprint{QMUL/05-2005}
\title{Chaos-order
  transition in Bianchi I non-Abelian Born-Infeld cosmology} \author{
  Vladimir V. Dyadichev} \email{vv_dyadichev@pisem.net} \author{
  Dmitri V. Gal'tsov}
\email{galtsov@mail.phys.msu.ru}\affiliation{Department of
Theoretical
  Physics, Moscow State University, 119899, Moscow, Russia}
\author{Paulo Vargas Moniz}
\email{p.moniz@qmul.ac.uk}
\homepage{http://webx.ubi.pt/~pmoniz}
\affiliation{Astronomy Unit, School of Mathematical Sciences,
University of London, Mile End Road, London E1 4NS, United
Kingdom\/}
%\altaffiliation[Also at~]{Centra--- IST, Rua
%  Rovisco Pais, 1049 Lisboa, Portugal}
\thanks{On leave of absence from
Departmento de Fisica, UBI,
  Covilha, Portugal;
Also at Centra---IST,  Lisboa,
  Portugal} 

\date{09.04.05}
\begin{abstract}
  We investigate the Bianchi I cosmology with the homogeneous SU(2)
  Yang-Mills field governed by the non-Abelian Born-Infeld action.
  A similar system with the standard Einstein-Yang-Mills (EYM) action is
  known to exhibit chaotic behavior induced by the Yang-Mills field.
  When the action is replaced by the Born-Infeld-type non-Abelian
  action (NBI), the chaos-order transition is observed in the high
  energy region. This is interpreted as a smothering effect due to
  (non-perturbative in $\alpha'$) string corrections to the classical
  EYM action. We give a numerical evidence for the chaos-order
  transition and present an analytical proof of regularity of color
  oscillations in the limit of strong Born-Infeld non-linearity. We
  also perform some general analysis of the Bianchi I NBI cosmology
  and derive an exact solution in the case when only the U(1)
  component of the Yang-Mills field is excited. Our new exact solution
  generalizes the Rosen solution of the Bianchi I Einstein-Maxwell
  cosmology to the U(1) Einstein-Born-Infeld theory.
\end{abstract}

\maketitle

\section{Introduction}

One of the key questions in theoretical cosmology is whether the
space-time metric near the singularity is regular or chaotic. As
was shown by Belinskii, Khalatnikov and Lifshitz (BKL), the
generic solution of the four-dimensional vacuum Einstein equations
exhibits an oscillating behavior \cite{BeKhLi70} which was later
qualified as essentially chaotic (see \cite{CoLe97} and references
therein). Recently the issue of chaos in the early universe
received a renewed attention due to discovery that the
antisymmetric form fields in ten and eleven-dimensional
supergravities imply chaos
\cite{DaHe00,DaHeJuNi01,DaHeReWe02,Damour:2002pi,Damour:2002fz}.
Namely, it was shown that the general solution near a space-like
singularity of the Einstein-dilaton-p-form field equations
exhibits an oscillatory behavior of the BKL type. However the
issue of chaos in superstring cosmology is not completely solved
yet, since these considerations are based on the lowest in
$\alpha'$ level of the string theory. To go beyond this
approximation in the closed string theories is difficult, since no
exact in $\alpha'$ effective action is known. Moreover, the
quadratic curvature corrections obtained perturbatively are likely
to disfavor chaos \cite{BaCo89, Coetal00}. Recently damping of the
BKL oscillations was observed also within the brane-world scenario
\cite{Co01}.

The role of non-perturbative string corrections is easier to study
within the open string context (or D-branes) where an exact in $\alpha'$
effective action is known in the slow varying field approximation under
the form of the Born-Infeld (BI) action \cite{FrTs85,Le89}.
A non-Abelian
generalization of the Born-Infeld action
was  suggested in ref. \cite{Ts99}.
Consequently, one can explore whether the Yang-Mills
(YM) chaos extensively
studied for homogeneous fields in flat space-time and in Bianchi
I cosmology persists when the usual YM action is replaced by the
non-Abelian Born-Infeld action (NBI). This would probe the effect of the
string non-locality on the issue of chaos.

Classical YM fields governed by the ordinary
quadratic action exhibit chaotic behavior in various situations
\cite{Bibook,Ma00}. The simplest case is that of the homogeneous
YM fields depending only on time in the flat space-time
\cite{BaMaSa79,MaSaTe81,ChSh81}: when only two YM components are
excited, the problem is reduced to the well-known two-dimensional
hyperbolic system $H=(p_x^2+p_y^2+x^2y^2)/2$ which is chaotic.
Furthermore, in
the lattice simulations of the inhomogeneous YM system, one
observes the energy flow from the infrared to the ultraviolet
region \cite{NiRuRu96}.
Therefore, it is believed that the chaotic behavior
is typical for the purely classical YM equations, one of the
arguments being the absence of solitons in this theory.
In addition,
it is known that adding the Higgs field to the YM theory leads to
stabilization of chaos in the homogeneous systems. In this case
the hyperbolic model is replaced by the system of coupled harmonic
oscillators which is regular in the weak coupling regime.

In the context of YM fields in the presence of gravity,
it is of interest to pint the following.
The YM field has violent oscillating behavior near the singularity
of the Einstein-Yang-Mills (EYM)
black holes \cite{DoGaZo97}, but the oscillations are not
chaotic. In the
domain of cosmology, some homogeneous  models, such as an
axisymmetric Bianchi I, the YM chaos unambiguously persists
\cite{DaKu96,DaKu97,BaLe98}, though in principle gravity leads to
a smoothing of the chaotic behavior. On the other hand, the Born-Infeld
effect on the flat-space dynamics of the homogeneous axisymmetric
YM field was shown to provide a chaos-order transition
\cite{DyGa03}, so it can be expected that in the gravity coupled
case this effect will be even more pronounced. Note that in the
related investigation of the behavior of the YM field inside
black holes \cite{DyGa00}, it was found that violent YM
oscillations disappear once the quadratic YM action is replaced by
the NBI action.

In this paper, we study in detail the axisymmetric Bianchi I cosmology with
the YM field governed by the NBI action. Though the system becomes
much more complicated when the ordinary YM Lagrangian is replaced
by the non-Abelian Born-Infeld (NBI) Lagrangian, we show that the
equations can be considerably simplified in some physically
interesting limiting cases, and even admit some exact solutions.
The generic solution exhibits a transition from the YM chaos   to
the regular oscillating regime when moving backward in time.

\section{General setting}

As was discussed recently \cite{Ha81,Ts97,Ts99,Pa99}, the definition
of the NBI action is ambiguous. One can start with the U(1) BI action
presented either in the determinant form
\begin{equation}\lb{det}
S=\frac1{16\pi}\int \,\sqrt{-\det(g_\mn-\beta^{-1}F_\mn)} \, d^4x,
\end{equation}
or in the equivalent (in four space-time dimensions) ``square root''
form
\begin{equation}\lb{sqrt} S=\frac1{16\pi} \int\, \sqrt{1+\frac{F_\mn
F^\mn}{2\beta^2}-\frac{({\tilde F}_\mn F^\mn )^2}{16\beta^4}}
\;\sqrt{-g} \, d^4x.
\end{equation}
In the non-Abelian case of the matrix-valued $F_\mn$, the trace over
gauge matrices must be specified. One particular definition is due to
Tseytlin~\cite{Ts97}. A symmetrized trace was therein introduced,
prescribing a symmetrization
of all products of $F_\mn$ in the power expansion of the determinant
(\ref{det}) before the trace is taken. Inside the symmetrized
series expansion the gauge generators effectively commute, so
both the determinant (\ref{det}) and the square root (\ref{sqrt})
forms are equivalent. This property does not hold for other trace
prescriptions, e.g., an ordinary trace. In the latter case it is
common to apply the trace to the square root form (\ref{sqrt}). Note
that string theory
seems to require the symmetrized trace definition in the
lower orders of the perturbation theory \cite{Ts97,HaTa97,Ts99},
while higher order corrections seem to violate this prescription
\cite{ReSaTe01,Bi01,Be01,SeTrTr01}. Here we choose the ``square
root/ordinary trace'' Lagrangian just for its simplicity. It is
worth noting that in the static case discussed recently both in the
ordinary~\cite{GaKe99} and the symmetrized trace~\cite{DyGa00} versions,
qualitative features of solutions turn out to be the same. Thus we
choose the action of the Einstein-NBI system in the following form
\begin{equation}\label{BItrace}
S=-\frac{1}{4\pi}\int \left\{\frac1{4G}R + \beta^2(\cR -
1)\right\}\;\sqrt{-g}\, d^4x,
\end{equation}
where $R$ is the scalar curvature, $\beta$ is the BI critical field
strength and
\begin{equation}\lb{cR}
\cR=\sqrt{1+\frac{1}{2\beta^2}F^a_\mn F_a^\mn -
\frac{1}{16\beta^4}(\duF^a_\mn F_a^\mn)^2}.
\end{equation}
The limit $\beta \to \infty$ corresponds to the standard
EYM theory with the action
\begin{equation}\label{YMtrace}
S=-\frac{1}{4\pi}\int \left(\frac1{4G}R + \frac14 F^a_\mn F_a^\mn
\right)\;\sqrt{-g}\, d^4x.
\end{equation}

We consider an axially symmetric Bianchi I space-time described by the
line element
\begin{equation}
 ds^2=N^2dt^2-b^2(dx^2+dy^2)-c^2dz^2,
\end{equation}
where functions $N$, $b$ and $c$ depend on time $t$. In the YM case this
problem was studied previously by Darian and Kunzle \cite{DaKu96,DaKu97}
and Barrow and Levin \cite{BaLe98}. The gauge field compatible with the
space-time symmetry is parameterized by two functions $u, v$ of time
\begin{equation}
A=T_1 u dx +T_2 u dy + T_3 v
  dz, \label{kunans}
\end{equation}
where SU(2) generators are normalized according to $[T_1, T_2]=i T_3$.
The corresponding field strength matrix-valued two-form is
\begin{align} & F=\dot{u}(T_1dt\wedge dx +T_2dt\wedge
  dy)+\dot{v}T_3 dt\wedge
  dz\nonumber\\
  &+u^2T_3 dx\wedge dy + uv(T_2 dz \wedge dx + T_1 dy \wedge
  dz).\label{kunf}
\end{align} Integrating over the 3-space, we obtain the
following one dimensional Lagrangian
\begin{equation} L=   - \frac12
\frac{\dot{b}(\dot{b}c+2\dot{c}b)}{GN}- \beta^2 N b^2 c (\cR - 1),
   \label{nbigrav}
\end{equation}
where now
\begin{align}
  &\cR=\sqrt{1-\frac{{\cal F}}{\beta^2} - \frac{{\cal G}^2}{\beta^4}}
,\label{sqrtdef}\\
  & {\cal F} = \frac{2\dot{u}^2}{N^2b^2}+ \frac{{\dot{v}}^2}{N^2 c^2}
  -\frac{1}{b^2}\left(\frac{2 u^2v^2}{  c^2 }+\frac{u^4 }{b^2}\right),\\
  & {\cal G} = \frac{u (2\dot{u}v+\dot{v}u)}{N b^2 c}.
\end{align}
The quantity $\cF$ is the YM Lagrangian, and it is
convenient to present it as a difference of kinetic and potential
terms
\begin{eqnarray} \cF&=&T-U,\\
T&=&\frac{2\dot{u}^2}{N^2b^2}+ \frac{{\dot{v}}^2}{N^2 c^2},\\
U&=& \left(\frac{2 u^2v^2}{ b^2 c^2 }+\frac{u^4 }{b^4}\right).
\end{eqnarray}
Note that from two coupling parameters entering the action, $G$ and
$\beta$, one can be eliminated by an appropriate rescaling. In what
follows we set $G=1$.

The Einstein equations can be derived by variation of the one-dimensional
action over $N, b, c$. Variation over $N$ gives the Hamiltonian
constraint
\begin{equation} \cH=\frac{\partial L}{\partial
N}=0,
\end{equation}
where $\cH$ reads in the synchronous gauge $N=1$:
\begin{equation}
\lb{H} \cH=\frac12
     \dot{b}(\dot{b}c+2\dot{c}b) + \frac{b^2 c}{\cR}
\left[\beta^2(\cR - 1)-U\right].
\end{equation}
Fixing this gauge from now on, we obtain the remaining Einstein
equations:
\begin{eqnarray}
\frac{\ddot {b}}{b}\!+\!\frac{\dot {b}}{b}\frac{\dot
{c}}{c}\!+\!\frac{\ddot {c}}{c}\!\!&=&\!\!2\beta^2(\cR\! -\!1)\!+\!\frac{2}{\cR}\!\Lef\!
\frac{\dot{u}^2}{b^2}\!-\!\frac{ u^2v^2}{ b^2 c^2 }\!-\!\frac{u^4
}{b^4}\!+\!\frac{\cG^2}{\beta^{2}}\!\Rig,\lb{E1}\\
\frac{{\ddot
b}}{b}+\frac12\frac{\dot {b}^2}{b^2} &=&\beta^2(\cR -
1)+\frac{1}{\cR}\Lef \frac{{\dot{v}}^2}{ c^2}-2\frac{ u^2v^2}{ b^2 c^2 }
+\frac{\cG^2}{\beta^2}\Rig.\lb{E2}
\end{eqnarray}

The equations for the YM field can be presented in the
following form
\begin{eqnarray}
\frac{\cR}{c}\frac{d}{dt}\left[\frac{c}{\cR}\Lef\dot{u}
+\frac{uv\cG}{c\beta^2}\Rig\right]\!\!&=&\frac{(\dot{u}v+\dot{v}u)\cG}{c\beta^2}\!\!-\frac{u^3}{b^2}-\frac{uv^2}{c^2},\\
 c\cR\!\frac{d}{dt}\left[\frac{b^2}{c\cR}\Lef\dot{v}
\!+\!\frac{cuv\cG}{ \beta^2}\Rig\right]\!\!&=&\!\!- 2u^2v
+\frac{2c\dot{u}u\cG}{ \beta^2}.
\end{eqnarray}

The energy-momentum tensor has the following components. The
energy density is given by
\begin{equation}
    T^0_0=\epsilon =
\frac{\beta^2+2\Psi^2\Gamma^2+\Psi^4}{4\pi\cR}-\frac{\beta^2
}{4\pi},
\end{equation}
the  pressure in the plane orthogonal to the symmetry axis is
\begin{equation}
p_x=-T^x_x=-T^y_y=\frac{\Pi_\Gamma^2+\Pi_\Psi^2-\Gamma^2\Psi
^2-\beta^2}{4\pi\cR}+
    \frac{\beta^2}{4\pi},
\end{equation}
and the  pressure along the axis of the symmetry is
\begin{equation}
    p_z=-T^z_z=\frac{2\Pi_\Psi-\Psi^4
-\beta^2}{4\pi\cR}+\frac{\beta^2}{4\pi}.
\end{equation}

\section{Reduction of order}

The above system of equations look as a dynamical system of the
eight-order in the presence of a constraint. However, it possesses
additional scaling
symmetries which can be used to reduce the system order by two (for
the EYM action this possibility was noticed by Darian
and Kunzle \cite{DaKu96}). It is easy to check that under a scaling
transformation
\[ b\to \lambda b,\quad
c\to \lambda^{-2} c,\quad u\to \lambda u,\quad v\to \lambda^{-2} v,
\]
the Lagrangian remains invariant. Moreover, under a separate rescaling
in the $b, u$ sector
\begin{equation}
b\to \lambda b, \quad u\to \lambda u,
\end{equation}
the Lagrangian scales as $\lambda^2$, and under the transformation
\begin{equation}c\to \lambda c,
\quad v\to \lambda v,
\end{equation}
as $\lambda$. The corresponding reduction of the EYM system is
achieved by an introduction of new variables invariant under the above
rescalings. Following Barrow and Levin \cite{BaLe98}, whose notation
we will adopt in what follows (note that in Ref. \cite{BaLe98} another
convention $8\pi G=1$ is used), we introduce the volume and shear
variables
\begin{equation}
 a=(b^2 c)^{1/3},\quad
\chi=\Lef\frac{b}{c}\Rig ^{1/3},
\end{equation}
together with the associated Hubble parameters
\begin{equation}
H_a= \frac{{\dot a}}{a},\quad H_\chi=
\frac{{\dot \chi}}{\chi},
\end{equation}
as well as the scaled Yang-Mills variables
\begin{equation} \1=\frac{u}{b},\quad \2=\frac{v}{c}.
\end{equation}
It is also convenient to use the scaled derivatives
\begin{equation}
\3=\frac{{\dot u}}{b},\quad
\4=\frac{{\dot v}}{c},
\end{equation}
which are related to ${\dot \1}, {\dot \2}$ via
\begin{eqnarray}
  &&\3={\dot \1}+\left(H_a+H_\chi\right)\1,\lb{W1}\\
  &&\4={\dot \2}+\left(H_a-2H_\chi\right)\2.\lb{W2}
 \end{eqnarray}
Note that these are not the momenta conjugate to $\1, \2$, the
 corresponding canonical momenta being
\begin{eqnarray}
P_\Psi&=&\frac{2a^3}{\cR}\Lef\3+\frac{\1\2\cG}{\beta^2}\Rig, \lb{pu}\\
P_\Gamma&=&\frac{a^3}{\cR}\Lef\4+\frac{\1^2\cG}{\beta^2}\Rig.\lb{pv}
\end{eqnarray}

In terms of the new variables the Hamiltonian constraint reads
\begin{equation}\label{Hnew}
 \frac32\left(H_a^2-H_\chi^2\right)+\beta^2-
\left[\beta^2+\1^2(\1^2+2\2^2)\right]\cR^{-1}=0 .
\end{equation}
The functions $T,U$ and $\cG$ entering $\cR$ now take the form
\begin{eqnarray}
T&=&2\3^2+\4^2,\lb{T}\\U&=&\1^4+2\1^2\2^2,\lb{U}\\
\cG&=&\1(2\3\2+\4\1).
\end{eqnarray}

From the Einstein equations one can derive two {\em first order}
equations for the Hubble parameters, which are linear in derivatives.
Taking the sum of the Eqs. (\ref{E1}),(\ref{E2}) and the constraint
equation (\ref{H}), one obtains the following simple equation for
$\dot{H}_a$:
\begin{equation}
\lb{1e}\dot{H}_a+3H_\chi^2+\frac{2}{3\cR}(T+U)=0.
\end{equation}
Similarly, taking twice the second Einstein equation (\ref{E2}) and
subtracting (\ref{E1}) we get
\begin{equation}
\lb{2e} \dot{H}_\chi+3H_\chi
H_a-\frac{2}{3\cR}(\4^2-\3^2+\1^4-\1^2\2^2)=0.
\end{equation}
Thus the Einstein equations reduce to two first order equations in the
presence of a constraint.

Alternatively, one can introduce the Hubble factors with respect to $b$
and $c$:
\begin{equation}
H_b= \frac{\dot{b}}{b},\quad H_c=
\frac{\dot{c}}{c},
\end{equation}
and bring  the Einstein equations into the form
\begin{eqnarray}
\dot{H}_b+H_b(H_b-H_c)&=&-\frac{2}{\cR}(\3^2+\1^2\2^2),\lb{E1a}\\
\dot{H}_c+H_c(H_b-H_c)&=&-\frac{2}{\cR}(\4^2+\1^4),\lb{E2a}
\end{eqnarray}
with the Hamiltonian constraint
\begin{equation}  \frac12H_b(H_b+2H_c)-
\frac1{\cR}\left[\beta^2(1-\cR)+U\right]=0.
\end{equation}

In addition, we have two second order equations for the YM
 fields which read in terms of the new variables
\begin{widetext}
 \begin{eqnarray}
\Lef \frac{d}{dt}+H_b+H_c\Rig\left[\frac1{\cR}
\Lef\3+\frac{\cG\1\2}{\beta^{2}}\Rig\right]+\frac1{\cR}
\left[\1^3+\1\2^2-\frac{\cG(\3\2+\4\1)}{\beta^{2}}\right]&=&0,\\
\Lef \frac{d}{dt}+2H_b\Rig\left[\frac1{\cR}
\Lef\4+\frac{\cG\1^2}{\beta^{2}}\Rig\right]+\frac2{\cR}
\left[\1^2\2+\1\2^2-\frac{\cG\3\1}{\beta^{2}}\right]&=&0.
\end{eqnarray}
\end{widetext}

\section{YM limit}

In the YM limit $\beta\to\infty$ the square-root factor in the
above formulas should be replaced according to the relation
\begin{equation}
 \lim_{\beta\to\infty}\beta^2(\cR-1) =-\frac12
\cF.\lb{ymlim}
\end{equation}
The main qualitative difference between EYM and ENBI theories lies in
the fact that the standard YM action is scale-invariant (though not the
EYM one) contrary to the NBI case.
This leads to a partial decoupling of
the YM dynamics from that of the space-time. Given eq.
(\ref{ymlim}), the
constraint equation simplifies to
\begin{equation}
 \frac12\left[3\left(H_a^2-H_\chi^2\right)-(T+U)\right]=0.
\end{equation}
Combining this with (\ref{1e}), one finds that one of the Einstein
equations fully decouples and reduces to the vacuum form:
\begin{equation}
\dot{H}_a+ H_\chi^2+2 H_a^2=0.
\end{equation}
However, the shear remains coupled to matter and obeys the equation
\begin{equation}
\dot{H}_\chi+3H_\chi H_a +H_a^2-H_\chi^2=\3^2+\1^4.
\end{equation}
Finally, the YM field equations become
\begin{eqnarray}
\dot{\Pi}_\Psi+(H_b+H_c)\3+\1(\1^2+\2^2)&=&0,\\
\dot{\Pi}_\Gamma+2H_b\4+ 2\1^2\2&=&0,
\end{eqnarray}
where the definitions (\ref{W1}), (\ref{W2}) have to be used.

The Hamiltonian form of the EYM equations can be further simplified
using an exponential parametrization of the volume and shear variables
\begin{equation}a=\e^\alpha,\quad
\chi=\e^\gamma.
\end{equation}The canonical momenta conjugate to $\alpha,
\gamma$ are
\begin{equation}
P_\alpha=-3\e^{3\alpha}\dot{\alpha},\quad
P_\gamma=3\e^{3\alpha}\dot{\gamma},
\end{equation}
while the YM momenta (\ref{pu}), (\ref{pv}) simplify to
\begin{equation}P_\Psi= 2a^3  \3 ,
 \quad P_\Gamma= a^3 \4.
\end{equation}
The Hamiltonian constraint (\ref{H}) for the EYM system in terms of the
momentum variables reads
\begin{equation}
\cH=\e^{-3\alpha}\left[ \frac16\Lef
P_\alpha^2-P_\gamma^2\Rig-\frac14\Lef
P_\Psi^2+2P_\Gamma^2\Rig\right]-\frac{U}{2}=0,
\end{equation}
where the potential is given by the Eq. (\ref{U}).

\section{U(1) case}

Consider the special case when only the $v$-component of the
YM field is excited, corresponding to the U(1) subgroup of
the gauge group. The Einstein equations (\ref{E1a}), (\ref{E2a})
reduce to
\begin{eqnarray}
\dot{H}_b+H_b(H_b-H_c)&=&0 ,\lb{E1v}\\
\dot{H}_c+H_c(H_b-H_c)&=&-\frac{2}{\cR}\4^2,\lb{E2v}
\end{eqnarray}
and the Hamiltonian constraint is
\begin{equation}
H_b(H_b+2H_c)=2\beta^2\Lef\cR^{-1}-1\Rig.\lb{Hv}
\end{equation} Integrating the BI
field equation
\begin{equation}
\frac{d}{dt}\Lef\frac{b^2\4}{\cR}\Rig=0,
\end{equation}
one obtains
\begin{equation}\frac{b^2\4}{\cR}=2b_0,\end{equation}
where $b_0$ is an integration constant, so that
\begin{equation}
\cR=\sqrt{1-\frac{\4^2}{\beta^2}}=\frac1{\sqrt{1+x^2}}, \quad
x=\frac{2b_0}{\beta b^2}.
\end{equation}
It is easy to see that the Einstein equation (\ref{E1v}) is equivalent
to
\begin{equation}\frac{\ddot{b}}{\dot{b}}=\frac{\dot{c}}{c},
\end{equation} which immediately gives a relation
\begin{equation}\dot{b}=kc,\lb{bc}\end{equation}where $k$ is a
second integration constant. Now the constraint equation becomes the
following separated equation for the function $b(t)$: \begin{equation}
  \dot{H}_b +\frac32 H_b^2=\beta^2\Lef\sqrt{1+x^2}-1\Rig, \lb{beq}
\end{equation}while the second Einstein equation (\ref{E2v}) is its time
derivative. The right hand side of this equation is positively
definite. It follows that the system has no bounces. Indeed, if
$H_b=0$, from the Eq. (\ref{E2v}) it follows that $\dot{H}_b=0$, which
contradicts the Eq. (\ref{beq}).

We can solve the Eq. (\ref{beq}) considering instead of $b(t)$ an
inverse function $t(b)$. Then
\begin{equation}H_b=\frac1{bt'},\end{equation}where $t'=dt/db$. The
equation for $t(b)$ following from (\ref{beq}) reads
\begin{equation}\Lef\frac1{t'}\Rig^2\Lef\frac{t''}{t'}-\frac1{2b}\Rig=
  b\beta^2\Lef1-\sqrt{1+\frac{4b_0^2}{\beta^2
      b^4}}\Rig.\end{equation}This is the linear first order equation
for the function \begin{equation}z(b)=
  (1/t')^2,\lb{z}\end{equation}namely,\begin{equation}
  z'+\frac{z}{b}+2b\beta^2\Lef1-\sqrt{1+\frac{4b_0^2}{\beta^2
      b^4}}\Rig =0.\end{equation}Its solution reads \begin{equation}
  z=\frac{2\beta^2}{b}\int\Lef \sqrt{1+\frac{4b_0^2}{\beta^2
      b^4}}-1\Rig b^2 db +\frac{b_1}{b},\end{equation}where $b_1$ is
a  third integration constant.
An integration can be done in terms of
the hypergeometric function \cite{AbSt}:
\begin{multline}z= \frac{2\beta^2}3
  \sqrt{b^4+\frac{4b_0^2}{\beta^2}}-\frac{2\beta^2b^2}3
+\\
\frac{8\beta b b_0}{3}
  F\Lef \frac13,\,\frac34;\,\frac54;\,\frac1{1+x^2}
\Rig+\frac{\tilde{b}_1}{b},
\end{multline}
where $\tilde{b}_1\neq b_1$ is another constant. Now, according to
(\ref{z}), the inverse function to the required solution is given by the
integral
\begin{equation}
  t(b)=\int\frac{db}{\sqrt{z(b)}}+t_0,\lb{tb}
\end{equation}where $t_0$
is the last integration constant in this
process. Our solution generalizes the Rosen
solution \cite{Ro62} to the Einstein-Born-Infeld theory.

Near the singularity $z\approx b_1/b$, so one has
\begin{equation}
  H_b=\frac{\sqrt{b_1}}{b^{3/2}}.
\end{equation}
Integrating Eq. (\ref{tb}) one obtains
\begin{equation}b=(b_1t)^{2/3},\end{equation}
and then from Eq. (\ref{bc})
\begin{equation}c= \frac{2b_1^{2/3}}{3k}
  t^{-1/3}.
\end{equation}
Hence, we obtain  a cigar singularity.

In the Maxwell case the situation is different. Indeed, in the limit
$\beta\to\infty$ one has
\begin{equation}z=-\frac{4b_0^2}{b^2}+
  \frac{b_1}{b}.
\end{equation}
Since $z$ should remain positive, the region of $b$ is limited by
\begin{equation}b>b_{{\rm min}}=\frac{4
    b_0^2}{b_1}.
\end{equation}
Combining the Eqs. (\ref{tb}), (\ref{bc}) we obtain
\begin{equation}b=b_{{\rm
min}}+\frac{b_1 t^2}{4
    b_{{\rm min}}},\quad c=\frac{b_1 t}{2 k b_{{\rm
min}}}.
\end{equation}
This is a pancake singularity. Thus, the BI non-linearity
modifies the singularity  from a pancake to a cigar type.

\section{Singularity structure}

Consider now the general solution near the cosmological
singularity. It turns out that except for a special isotropic
solution $b=c=a$, previously studied in
\cite{DyGaZoZo01,Mo02,Mo02q,Mo02f}, generic solutions have the
same metric singularities as the vacuum Bianchi I solutions. Near
the pancake singularity the solution is not analytic in terms of
$t$, but in terms of $t^{1/3}$. In fact, one finds the following Laurent
expansion containing four free parameters $\x, \y, \z, \w$:
\begin{eqnarray}
H_a&=&\frac{1}{3t}-\frac{2\z\y+\w\x}{9\x\z}\,t^{-2/3}+O(t^{-1/3}),\\
H_\chi&=&\frac{1}{3t}+\left(\frac{\x \w-\y \z}{9
\x\z}-\frac{\x}{\sqrt{2}}\right)\,t^{-2/3} +O(t^{
-1/3 } ) ,\\
  \Psi&=&\x\,t^{-2/3}+\y\,t^{-1/3}+O(1),\\
  \Gamma&=&\z\,t^{1/3}+\w\,t^{2/3}+O(t).
\end{eqnarray}
The $\2$-component of the YM field vanishes at $t=0$, while $\1$ is
singular. The scale factor $a$ and the shear $\chi$ near the pancake
singularity both behave as $O(t^{1/3})$.

Near the cigar singularity the solution has a Laurent expansion in terms
of $t$:
\begin{eqnarray}\label{cig-expans}
  H_a&=&\frac{1}{3t}+\frac{4 \bar{\z}^2 \bar{\x}^2-\bar{\w}^2}{3 \cR_1}+ O(t),\\
  H_{\chi}&=&-\frac{1}{3t}+\frac{2 \bar{\z}^2
\bar{\x}^2+\bar{\w}^2}{3\cR_1}+O(t),\\
  \Psi&=& \bar{\x} + \left(\bar{\y}-\frac{2\bar{\z}^2
\bar{\x}^3}{\cR_1}\right)\,t + O(t^2),\\
%  \Pi_\Psi&=& y+ O(t) \\
  \Gamma&=& \bar{\z}\,t^{-1} + \bar{\w}+\frac{\bar{\z}\bar{\w}^2}{\cR_1} +O(t) ,
%  \Pi_\Gamma&=& w\,t^{-1} + O(1)
\end{eqnarray}
where the quantity $\cR_1$ is the leading term in an expansion of the
NBI square root:
\begin{equation}\cR=\cR_1 \,t^{-1}+O(1),\quad \cR_1 =
\sqrt{\frac{2 \bar{\z}^2 \bar{\x}^2-\bar{\w}^2}{\beta^2}-\frac{\bar{\x}^2
(\bar{\x} \bar{\w}+2 \bar{\z}
    \bar{\y})^2}{\beta^4}}.
\end{equation}
The scale factor and the shear have the following expansions:
\begin{eqnarray}
 a&=&a_1\left(t^{1/3}+\frac{4 \bar{\z}^2 \bar{\x}^2-\bar{\w}^2}{3 \cR_1}\,
        t^{4/3} +O(t^{7/3})\right),\\
\chi&=&\chi_1\left(t^{-1/3}+\frac{2 \bar{\z}^2
\bar{\x}^2+\bar{\w}^2}{3\cR_1}\,t^{2/3}+O(t^{5/3})\right).
\end{eqnarray}

The quantities $\x$, $\y$, $\z$, $\w$ ($\bar\x$, $\bar\y$, $\bar\z$,
$\bar\w$) are independent free parameters
which, together with an arbitrariness associated with a time shift,
provide five constants needed to specify the generic solution for both
singularity types.

\section{Solution in the limit $\beta=0$}

In order to better understand the effect of the BI
nonlinearities on the gauge field dynamics let us first study the
strong field limit $F\gg \beta$, or, formally, $\beta\to 0$. The
leading term in the square root (\ref{sqrtdef}) containing the
pseudoscalar invariant $\cG$ is negative definite. Therefore,
imposing the square root $\cR$ to be
real-valued in the limit $\beta\to 0$ may be
ensured only if $\cG$ tends to zero, in which case
\begin{equation}\lb{cons1}
    \Psi\Pi_\Gamma+2\,\Gamma\Pi_\Psi=0.
\end{equation}
One can show that this condition is compatible indeed with the
equations of motion as $\beta\to 0$.

Given the condition (\ref{cons1}), the square root term will read
\begin{equation}\label{root}
\mathcal{R}={\frac{\sqrt{\Psi^{4}+2\,\Gamma^{2}\Psi^{2}-
\Pi_\Gamma^{2}-2\,\Pi_\Psi^{2}}}{\beta}}.
\end{equation}
The right hand sides of the Einstein equations (\ref{1e}), (\ref{2e})
tend to zero, so the gravitational degrees of freedom decouple
\begin{equation}
  \dot{H}_a=-3\, H_\chi ^{2},\quad \dot{H}_\chi=-3\,H_\chi
  H_a,
\end{equation}
and the gravitational constraint assumes the vacuum form as well
\begin{equation} H_a^{2}- H_\chi^{2}=0. \end{equation}

Decoupling of gravity means that in the limit $\beta\to 0$ the metric is
given by the vacuum Kasner solution either of a cigar type
\begin{equation}
  H_a=H_\chi=\frac1{3t},
\end{equation}
or a pancake type
\begin{equation}   H_a=-H_\chi=\frac1{3t},\end{equation}
where we set the singularity at $t=0$.

Substituting the explicit expressions for the Hubble and shear
parameters   one  finds another  constraint
\begin{equation}\label{cons2}
    {\frac{\Psi^{2}\Gamma}{H_a}}=C=\mbox{const},
\end{equation}
and thus in the remaining equations one can express all the gauge field
variables either in terms of $\Gamma,\, \Pi_\Gamma$, or in terms of
$\Psi,\, \Pi_\Psi$. One simple consequence of this constraint is that in
the non-trivial case $C\neq 0$ the variables $\Psi$ and $\Gamma$ can not
have zeroes except for the singularity, and thus should preserve their
signs. From the NBI field equations one then finds
\begin{eqnarray}\Pi_\Psi&=&\dot{\Psi},\qquad \mbox{cigar},\\
  \Pi_\Psi&=&\dot{\Psi}-\frac{2\Psi}{3t},\qquad
  \mbox{pancake}.\end{eqnarray}
 In both cases the dynamical equation for $\1$ will be of the
 form
\begin{equation}\ddot{\Psi}=f(\Psi, \dot{\Psi}, C, t)\end{equation}
with some function $f$. It describes oscillations with a decreasing
amplitude. The second YM variable $\2$ is related to $\1$ algebraically
via constraints (\ref{cons1}), (\ref{cons2}) and therefore oscillates
with the same frequency exactly in an antiphase. Oscillations are
fully regular, so no YM chaos can persist in the regime of the strong
BI non-linearity.

The general solution near the pancake singularity can be expanded
with respect to the variable $\tau\equiv t^{1/3}$:
\begin{eqnarray}
\Gamma = p_1^2\,\tau+\sqrt{6C}p_1\,\tau^2+\frac{3C}{2} \tau^3+q_1 \tau^4+O(x^5),\\
%\Pi_{\Gamma} = \pm\sqrt{\frac{2pC}{3}}\,x^{-1}+C+q \,x\\
\Psi
=\frac{\sqrt{C}}{\sqrt{3}p_1}\,\tau^{-2}+\frac{C}{\sqrt{2}p_1^2}\,
\tau^{-1}+\frac{\sqrt{3}C^{3/2}}{2 p_1^3} + O(\tau),
\end{eqnarray}
where $p_1$ and $q_1$ are free parameters.

Near the cigar singularity the solution can be expanded in terms of $t$:
\begin{eqnarray}
\Gamma = p_1\,t^{-1}+q_1+\frac{3 q_1^2-p_1 C}{6 p_1}\,t+ O(t^2),\\
\Psi =
\frac{\sqrt{C}}{\sqrt{3p_1}}-\frac{\sqrt{C}q_1}{2\sqrt{3}p_1^{3/2}}\,t
+ O(t^2).
\end{eqnarray}

\section{Chaos-order transition}

Now we address the problem numerically. Various methods were suggested
to study a chaotic behavior in the context of gravity,
 where the absence of
the canonical time variable
prevents a straightforward use  of such convenient
tools as the Lyapunov exponents (however, see \cite{Imponente}). In
the case of the conformally invariant YM Lagrangian, one can
use the approach of Ref. \cite{BaLe98} to separate the dynamics of
the YM field from the gravitational expansion and then apply
the invariant technique of chaotic scattering. For systems exhibiting
chaotic behavior the set of all periodic orbits has fractal structure
invariant under coordinate reparameterizations.

However these methods become problematic in our case,
where the  conformal
invariance is absent from the
 matter action. Apart from the special asymptotic
regimes, it is not possible to separate the YM dynamics from the metric
evolution. Though it can be done for the high YM intensity, it turns out
that the time interval in the actual evolution where this regime holds
is rather small for $\beta$ of the order of unity or greater. The YM
variables perform only a small number of oscillations during the epoch
of high field intensity, the field then being fast diluted. However, if
we set the parameter $\beta$ sufficiently small, the time spent by the
system in the highly nonlinear region will be large enough, and in this
case the chaos-order transition is unambiguously manifest.

A numerical analysis of the system for small values of $\beta$ reveals
the following. While the gauge field strength is considerably greater
then the critical field $\beta$, both conditions (\ref{cons1}) and
(\ref{cons2}) approximately hold, and the dynamics of the gauge field
qualitatively coincides with that discussed in the previous section
for $\beta=0$. Both variables $\Psi$ and $\Gamma$ perform nonlinear
oscillations with decreasing amplitude in an antiphase with respect to
each other and without crossing zero. In this region the dynamics is fully
regular, while the evolution of the metric is mainly governed by purely
vacuum terms and is close to the vacuum Kasner solution.

\begin{figure*}
\centerline{\includegraphics{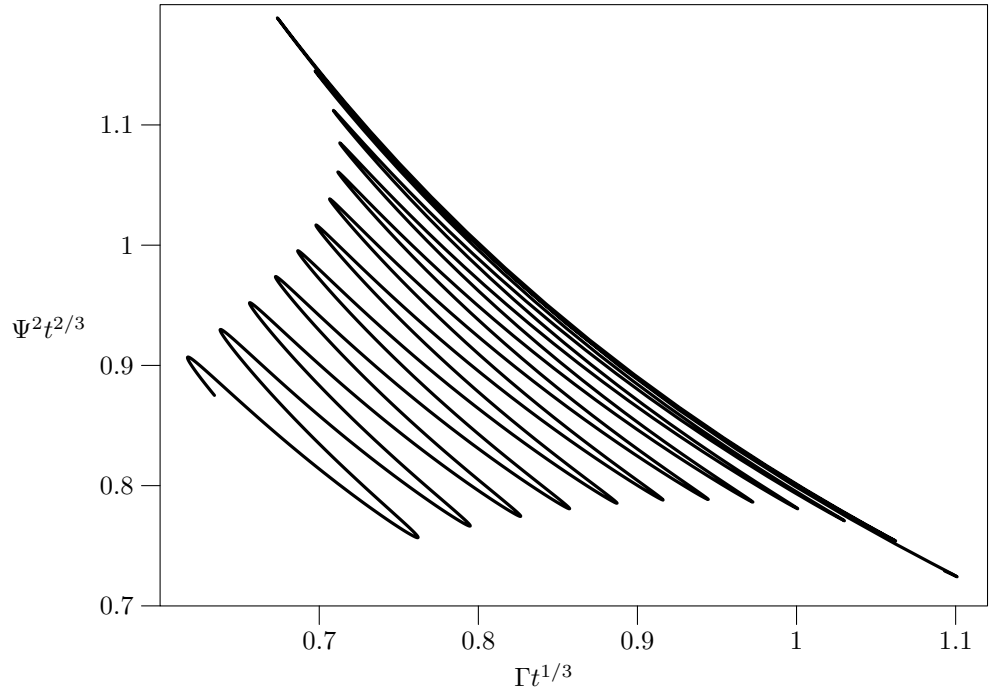}} \caption{The phase portrait
$\Psi^2 t^{2/3}$ vs. $\Gamma t^{1/3}$ for $\beta=10^{-4}$.}
\label{fig:constraint}
\end{figure*}

To test the validity of an approximate description of the system
in terms of the limiting $\beta=0$ solution we check the
constraint equations (\ref{cons1},\ref{cons2}) in the case of
small but finite $\beta$. The first constraint is the condition of
smallness of the pseudoscalar YM invariant $\cG^2\ll \beta^2 \cF
$. The second constraint was directly checked numerically.
Fig.~\ref{fig:constraint} illustrates the situation for the cigar
solution with $\beta=10^{-4}$. From this figure one can see that
the right hand side of Eq. (\ref{cons2}) evolves on time scales
much larger than the period of oscillations of the gauge field
variables. Under the overall volume expansion, the YM energy
density falls down and the role of the BI nonlinearity decreases.
At the same time the matter terms in the Einstein equation become
more significant forcing, in particular, to decrease the shear
anisotropy $H_{\chi}$ much faster than the Hubble parameter $H_a$.

\begin{figure*}
  \centerline{\includegraphics{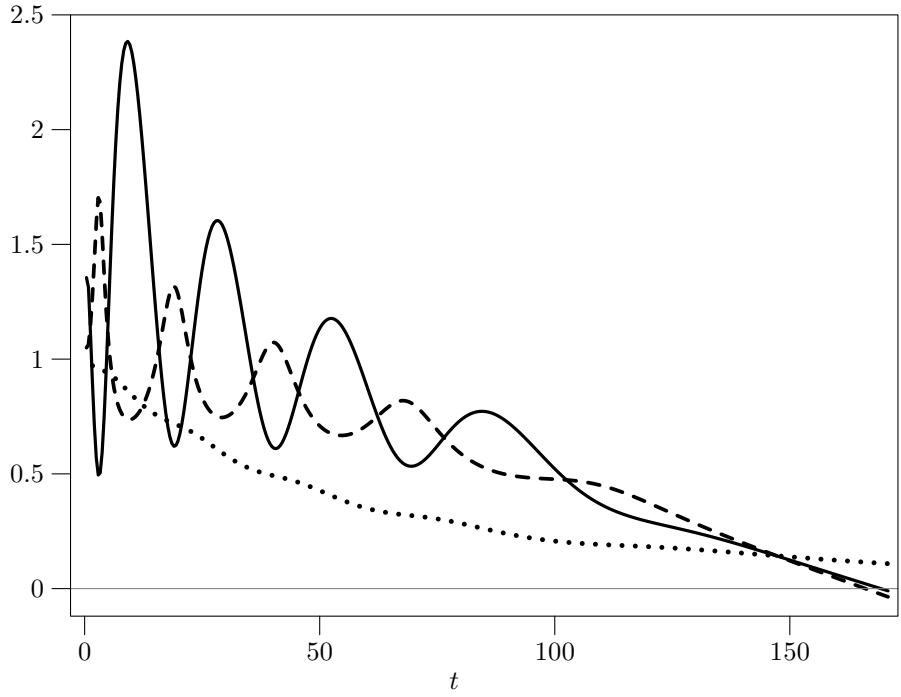}}
\caption{The solution for the $\beta=2\cdot10^{-3}$, regular
phase, a  cigar singularity. The solid line~--- $\Gamma t^{1/3}$,
  the dashed line~--- $\Psi t^{1/3}$, the dotted line~---
  $H_{\chi}/H_{a}$.}
\label{fig:reg}
\end{figure*}

\begin{figure*}
  \centerline{\includegraphics{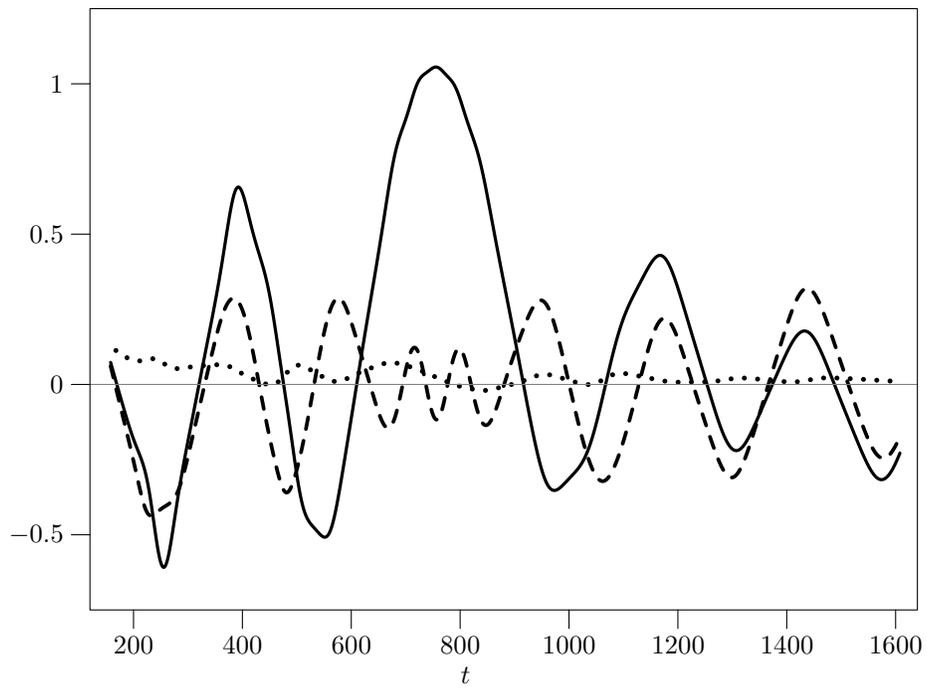}}
\caption{Further development of the solution from
  Fig.~\ref{fig:reg}~--- chaotic oscillations.}
\label{fig:cha}
\end{figure*}

\begin{figure*}[t]
  \centerline{\includegraphics{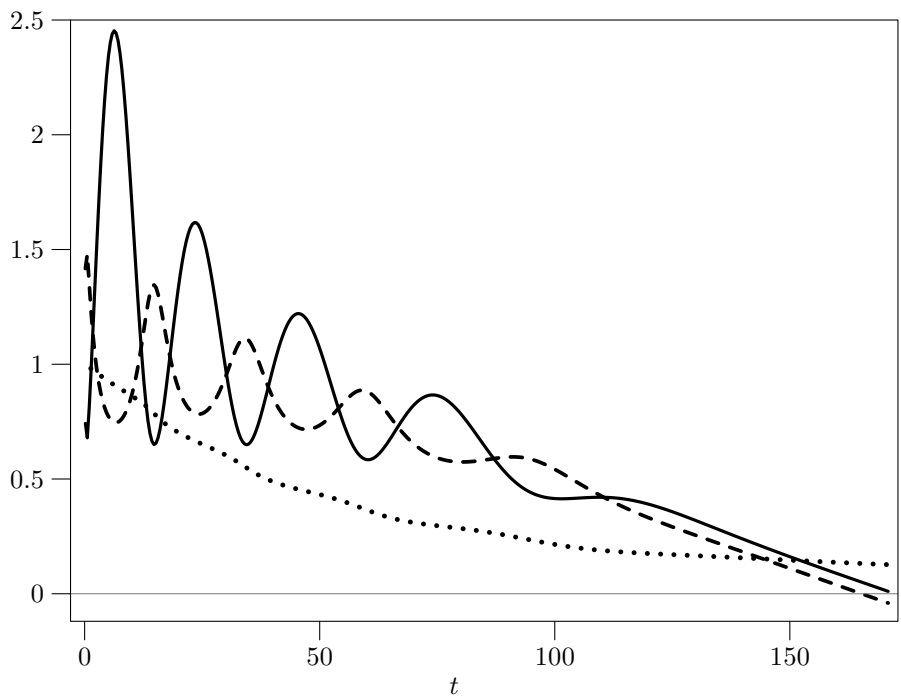}}
  \caption{The solution for $\beta=2\cdot10^{-3}$, a regular phase, a
pancake singularity. The solid line~--- $\Gamma t^{1/3}$, the dashed
line~--- $\Psi  t^{1/3}$, the dotted line~--- $-H_{\chi}/H_{a}$.}
\label{fig:pan}
\end{figure*}

These features are illustrated in
Figs.~\ref{fig:reg}--\ref{fig:pan}.  Fig.~\ref{fig:reg} shows the
early regular evolution for $\beta=2\cdot 10^{-3}$ and the cigar-type
singularity. Both variables $\Psi$ and $\Gamma$ oscillate in the
positive region. The behavior of the shear anisotropy $H_{\chi}$ is
smooth.  Fig.~\ref{fig:cha} demonstrates the same solution at
late time. One can see that the dynamics of the gauge field becomes
essentially chaotic. The function $H_{\chi}$ coupled to matter performs
chaotic oscillations with decreasing (as compared to the Hubble
parameter $H_a$) amplitude. The first zeroes of the gauge functions
$\Psi$ and $\Gamma$ serve as an approximate boundary separating highly
nonlinear evolution from the region of the chaotic regime. The actual
type of singularity (pancake or cigar), plays a relatively small role in
both the regular and strongly chaotic phases except for the small vicinity
of the singularity. This is illustrated in Fig.~\ref{fig:pan} which
shows the solution with pancake singularity obtained from the solution
shown in Fig.~\ref{fig:reg} by changing the sign of $H_\chi$ with initial
conditions which were set at $t=10$.

\begin{figure*}
  \centerline{\includegraphics{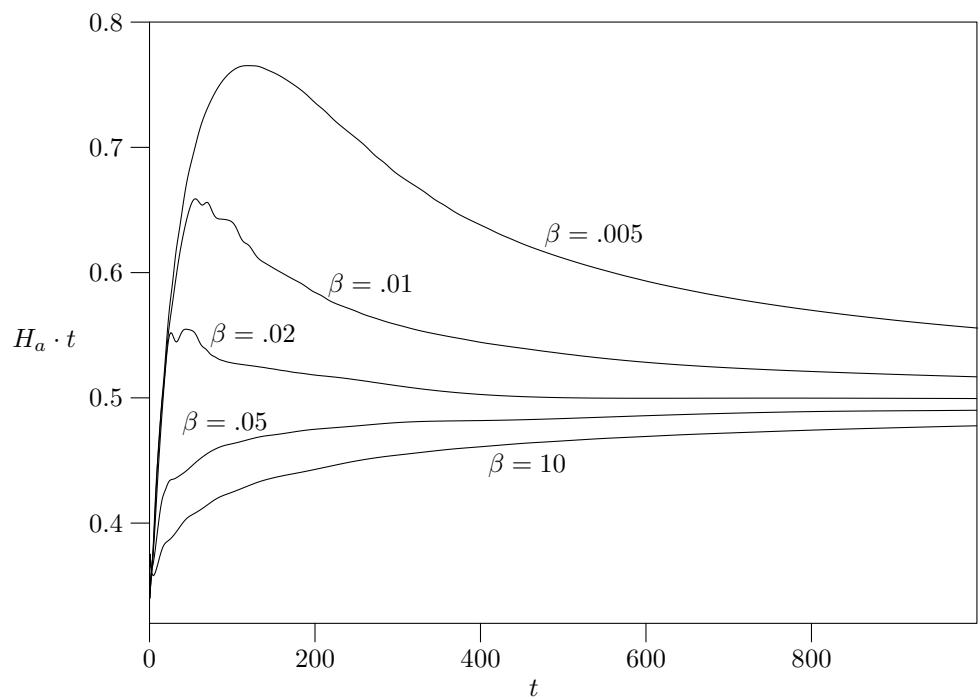}}
  \caption{The behaviour of
    $h(t)= t H_a$ for various values of $\beta$.}
  \label{fig:hubble}
\end{figure*}

The Hubble parameter $H_{a}$ does not exhibit chaotic behavior. It
can be presented as $H_a=h(t) t^{-1}$, with some slowly varying
smooth function $h(t)$. Numerical curves $h(t)$ are shown in
Fig.~\ref{fig:hubble} for various $\beta$. This function interpolates
between the value $1/3$ at $t=0$ (vacuum Kasner solution) and the value
$1/2$ at $t=\infty$ corresponding to the isotropic ``hot universe''
cosmology. However, for small values of $\beta$, when the system
stays in a highly nonlinear regime for a considerable time interval,
there is a region where $h(t)$ is greater than $1/2$. This feature
can be explained using the results of the FRW-BI model
and was firstly presented in ref.
\cite{DyGaZoZo01}
(Subsequent analysis can
be found in ref. \cite{Mo02,Mo02q,Mo02f}.).
 Eq. \ref{Hnew} implies that once the contribution
of the anisotropy term $H_\chi$ decreases (i.e. the solution undergoes
isotropisation), the Hamiltonian constraint tends to the Friedmann
equation. In the FRW case \cite{DyGaZoZo01} one can derive an equation
of state for  the NBI matter which interpolates between that for conformal
matter $\epsilon=1/3p$ and the ``string fluid'' equation $\epsilon =
-1/3p$ in the highly nonlinear regime. The latter corresponds to the
value $h=1$, which is, however, never achieved in the anisotropic case
which we investigate here.

%Behavior of the  Hubble parameter $H_{a}$ is regular for all
%times. It can be described by $H_a=F(t) t^{-1}$, with some slowly
%varying smooth function $F(t)$  interpolating between the values
%$1/3$ at $t=0$ and $1/2$ at $t=\infty$. Fig.~\ref{fig:hubble}
%demonstrates its behavior for various values of  parameters.

\section{Discussion}
The main goal of this paper was to test the non-perturbative effects
of superstring theory on the issue of chaos in cosmology. At
least three different patterns of chaotic
behavior in cosmology were identified. The
first is the billiard-type behavior which is
manifest in the Bianchi IX pure gravity and its supergravity (including
multidimensional cases) generalizations. The second is the bouncing
behavior of the FRW-scalar field cosmology. The third type is the
matter-dominated chaos of the Bianchi I EYM cosmology,
and it is this type of chaos which was investigated here (Recently
an interesting analysis was performed \cite{JiMa04} of the YM field
behavior in more general type A Bianchi space-times showing that basic
features of the YM chaos persist there as well). From these three
patterns
the last one is the most appropriate for testing the superstring
non-locality effects accumulated in the BI non-Abelian action.
Our results clearly demonstrate disappearance of chaos in the high
energy regime.

From a mathematical viewpoint, it is worth noting that the Einstein-NBI
system of equations admits a reduction of order due to
the presence of
scaling symmetries
similarly to the EYM case. Moreover, in the strong
BI regime the axisymmetric Bianchi I NBI system can be reduced
further due to existence of two additional asymptotic integrals of
motion. This limit is characterized by the dynamical vanishing of the
pseudoscalar quadratic invariant of the YM field. This simplifies
dynamics considerably and leads to a decoupling of the gravitational
degrees of freedom. Color oscillations are still governed by the
BI non-linearity and are reducible to the one-variable second
order system predicting perfectly regular behavior.

Numerical experiments shows that the system behavior for sufficiently
small $\beta$ consists of a regular phase in the high energy region near
the singularity and the chaotic phase at later time. The regular phase
is qualitatively similar to that described by the $\beta=0$ approximate
description. The chaos-order transition is observed when one is moving
backward in time towards the singularity. The singularity itself is
either of a cigar or a pancake type, as in the vacuum Bianchi I case,
though the YM field does not tend to the vacuum configuration. Thus
the non-perturbative in $\alpha'$ string corrections to the YM action
suppress the YM chaos which takes place at lower energies where dynamics
of the YM field is governed by the ordinary quadratic action.

In the case of only an Abelian component excited, an exact analytic
solution of the Einstein-BI system was found which generalizes
the Rosen solution to the Einstein-Maxwell equations. It also
exhibits a different behavior in the singularity as compared with the
Einstein-Maxwell case.

\begin{acknowledgments}
  D.V.G. thanks the University of Beira Interior for hospitality and
the grant NATO CR(RU)05/C/03 PO - GRICES - 03675 for support in
the summer 2004, when this research was
  initiated.  This work was also supported in part by the RFBR grant
  02-04-16949, POCTI(FEDER) P - FIS - 57547/2004 and
  CERN P - FIS - 49529/2003. PVM is supported by the grant
   FCT (FEDER) SFRH - BSAB 396/2003. He also  thanks QMUL for hospitality and
  R. Tavakol for conversations.
\end{acknowledgments}

\nocite{Ro62}

%\bibliography{chaos} \bibliographystyle{hunsrt}

\newpage

\end{document}